\documentclass{revtex4}

\usepackage{latexsym}
\usepackage{graphicx}

\newcommand{\nc}{\newcommand}

\nc{\be}[1]{\begin{equation}\mbox{$\label{#1}$}}
\nc{\bea}[1]{\begin{eqnarray} \mbox{$\label{#1}$}}
\nc{\Section}[2]{\section{#2}\label{#1}}
\nc{\Bibitem}[1]{\bibitem{#1}}
\nc{\Label}[1]{\label{#1}}

\nc{\eea}{\end{eqnarray}}
\nc{\ee}{\end{equation}}

\nc{\bdm}{\begin{displaymath}}
\nc{\edm}{\end{displaymath}}
\nc{\dpsty}{\displaystyle}
\nc{\bc}{\begin{center}}
\nc{\ec}{\end{center}}
\nc{\ba}{\begin{array}}
\nc{\ea}{\end{array}}
\nc{\bab}{\begin{abstract}}
\nc{\eab}{\end{abstract}}
\nc{\btab}{\begin{tabular}}
\nc{\etab}{\end{tabular}}
\nc{\bit}{\begin{itemize}}
\nc{\eit}{\end{itemize}}
\nc{\ben}{\begin{enumerate}}
\nc{\een}{\end{enumerate}}
\nc{\bfig}{\begin{figure}}
\nc{\efig}{\end{figure}}

\nc{\arreq}{&\!=\!&}
\nc{\arrmi}{&\!-\!&}
\nc{\arrpl}{&\!+\!&}
\nc{\arrap}{&\!\!\!\approx\!\!\!&}
\nc{\non}{\nonumber}

\def\lsim{\; \raise0.3ex\hbox{$<$\kern-0.75em
      \raise-1.1ex\hbox{$\sim$}}\; }
\def\gsim{\; \raise0.3ex\hbox{$>$\kern-0.75em
      \raise-1.1ex\hbox{$\sim$}}\; }

\nc{\DOT}{\hspace{-0.08in}{\bf .}\hspace{0.1in}}
\nc{\Laada}{\hbox {$\sqcap$ \kern -1em $\sqcup$}}
\nc\loota{{\scriptstyle\sqcap\kern-0.55em\hbox{$\scriptstyle\sqcup$}}}
\nc\Loota{{\sqcap\kern-0.65em\hbox{$\sqcup$}}}
\nc\laada{\Loota}
\nc{\qed}{\hskip 3em \hbox{\BOX} \vskip 2ex}

\nc{\real}{{\rm I \! R}}
\nc{\Z}{{\sf Z \!\!\! Z}}
\nc{\complex}{{\rm C\!\!\! {\sf I}\,\,}}
\def\bigid{\leavevmode\hbox{\small1\kern-3.8pt\normalsize1}}
\def\id{\leavevmode\hbox{\small1\kern-3.3pt\normalsize1}}
\nc{\slask}{\!\!\!/}
\nc{\bis}{{\prime\prime}}
\nc{\pa}{\partial}
\nc{\na}{\nabla}
\nc{\ra}{\rangle}
\nc{\la}{\langle}
\nc{\goto}{\rightarrow}
\nc{\swap}{\leftrightarrow}

\nc{\EE}[1]{ \mbox{$\cdot10^{#1}$} }
\nc{\abs}[1]{\left|#1\right|}
\nc{\at}[2]{\left.#1\right|_{#2}}
\nc{\norm}[1]{\|#1\|}
\nc{\abscut}[2]{\Abs{#1}_{\scriptscriptstyle#2}}
\nc{\vek}[1]{{\rm\bf #1}}
\nc{\integral}[2]{\int\limits_{#1}^{#2}}
\nc{\inv}[1]{\frac{1}{#1}}
\nc{\dd}[2]{{{\partial #1}\over{\partial #2}}}
\nc{\ddd}[2]{{{{\partial}^2 #1}\over{\partial {#2}^2}}}
\nc{\dddd}[3]{{{{\partial}^2 #1}\over
    {\partial #2 \partial #3}}}
\nc{\dder}[2]{{{d #1}\over{d #2}}}
\nc{\ddder}[2]{{{d^2 #1}\over{d {#2}^2}}}
\nc{\dddder}[3]{{d^2 #1}\over
    {d #2 d #3}}
\nc{\dx}[1]{d\,^{#1}x}
\nc{\dy}[1]{d\,^{#1}y}
\nc{\dz}[1]{d\,^{#1}z}
\nc{\dl}[1]{\frac{d\,^{#1}l}{(2\pi)^{#1}}}
\nc{\dk}[1]{\frac{d\,^{#1}k}{(2\pi)^{#1}}}
\nc{\dq}[1]{\frac{d\,^{#1}q}{(2\pi)^{#1}}}

\nc{\bfT}{{\bf T }}

\nc{\cA}{{\cal A}}
\nc{\cB}{{\cal B}}
\nc{\cD}{{\cal D}}
\nc{\cE}{{\cal E}}
\nc{\cG}{{\cal G}}
\nc{\cH}{{\cal H}}
\nc{\cL}{{\cal L}}
\nc{\cO}{{\cal O}}
\nc{\cT}{{\cal T}}
\nc{\cN}{{\cal N}}
\nc{\cR}{{\cal R}}
\nc{\rvac}[1]{|{\cal O}#1\rangle}
\nc{\lvac}[1]{\langle{\cal O}#1|}
\nc{\rvacb}[1]{|{\cal O}_\beta #1\rangle}
\nc{\lvacb}[1]{\langle{\cal O}_\beta #1 |}
\nc{\bb}{\bar{\beta}}
\nc{\bt}{\tilde{\beta}}
\nc{\ctH}{\tilde{\cal H}}
\nc{\chH}{\hat{\cal H}}
\nc{\1}{\aa}
\nc{\2}{\"{a}}
\nc{\3}{\"{o}}
\nc{\4}{\AA}
\nc{\5}{\"{A}}
\nc{\6}{\"{O}}
\nc{\al}{\alpha}
\nc{\g}{\gamma}
\nc{\Del}{\Delta}
\nc{\e}{\textrm{e}}
\nc{\eps}{\epsilon}
\nc{\lam}{\lambda}
\nc{\Om}{\Omega}
\nc{\ve}{\varepsilon}
\nc{\mn}{{\mu\nu}}
\nc{\vp}{\varphi}

\nc{\advp}[3]{{\it  Adv.\ in\ Phys.\ }{{\bf #1} {(#2)} {#3}}}
\nc{\annp}[3]{{\it  Ann.\ Phys.\ (N.Y.)\ }{{\bf #1} {(#2)} {#3}}}
\nc{\apjj}[3]{{\it  Ap.\ J.\ }{{\bf #1} {(#2)} {#3}}}
\nc{\apjl}[3]{{\it  Ap.\ J.\ Lett.\ }{{\bf #1} {(#2)} {#3}}}
\nc{\app}[3]{{\it Astropart.\ Phys.\ }{{\bf #1} {(#2)} {#3}}}
\nc{\cmp}[3]{{\it  Comm.\ Math.\ Phys.\ }{{ \bf #1} {(#2)} {#3}}}
\nc{\cqg}[3]{{\it  Class.\ Quant.\ Grav.\ }{{\bf #1} {(#2)} {#3}}}
\nc{\epl}[3]{{\it  Europhys.\ Lett.\ }{{\bf #1} {(#2)} {#3}}}
\nc{\ijmp}[3]{{\it Int.\ J.\ Mod.\ Phys.\ }{{\bf #1} {(#2)} {#3}}}
\nc{\ijtp}[3]{{\it Int.\ J.\ Theor.\ Phys.\ }{{\bf #1} {(#2)} {#3}}}
\nc{\jmp}[3]{{\it  J.\ Math.\ Phys.\ }{{ \bf #1} {(#2)} {#3}}}
\nc{\jpa}[3]{{\it  J.\ Phys.\ A\ }{{\bf #1} {(#2)} {#3}}}
\nc{\jpc}[3]{{\it  J.\ Phys.\ C\ }{{\bf #1} {(#2)} {#3}}}
\nc{\jap}[3]{{\it J.\ Appl.\ Phys.\ }{{\bf #1} {(#2)} {#3}}}
\nc{\jpsj}[3]{{\it J.\ Phys.\ Soc.\ Japan\ }{{\bf #1} {(#2)} {#3}}}
\nc{\lmp}[3]{{\it Lett.\ Math.\ Phys.\ }{{\bf #1} {(#2)} {#3}}}
\nc{\mpl}[3]{{\it  Mod.\ Phys.\ Lett.\ }{{\bf #1} {(#2)} {#3}}}
\nc{\ncim}[3]{{\it  Nuov.\ Cim.\ }{{\bf #1} {(#2)} {#3}}}
\nc{\np}[3]{{\it  Nucl.\ Phys.\ }{{\bf #1} {(#2)} {#3}}}
\nc{\pr}[3]{{\it Phys.\ Rev.\ }{{\bf #1} {(#2)} {#3}}}
\nc{\prll}[3]{{\it Phys\ Rev.\ Lett.\ }{{\bf #1} {(#2)} {#3}}}
\nc{\pll}[3]{{\it  Phys.\ Lett.\ }{{\bf #1} {(#2)} {#3}}}
\nc{\prep}[3]{{\it Phys\. Rep.\ }{{\bf #1} {(#2)} {#3}}}
\nc{\prsl}[3]{{\it Proc.\ R.\ Soc.\ London\ }{{\bf #1} {(#2)} {#3}}}
\nc{\ptp}[3]{{\it  Prog.\ Theor.\ Phys.\ }{{\bf #1} {(#2)} {#3}}}
\nc{\ptps}[3]{{\it  Prog\ Theor.\ Phys.\ suppl.\ }{{\bf #1} {(#2)} {#3}}}
\nc{\physa}[3]{{\it  Physica\ A\ }{{\bf #1} {(#2)} {#3}}}
\nc{\physb}[3]{{\it  Physica\ B\ }{{\bf #1} {(#2)} {#3}}}
\nc{\phys}[3]{{\it Physica\ }{{\bf #1} {(#2)} {#3}}}
\nc{\rmpp}[3]{{\it  Rev.\ Mod.\ Phys.\ }{{\bf #1} {(#2)} {#3}}}
\nc{\rpp}[3]{{\it Rep.\ Prog.\ Phys.\ }{{\bf #1} {(#2)} {#3}}}
\nc{\sjnp}[3]{{\it Sov.\ J.\ Nucl.\ Phys.\ }{{\bf #1} {(#2)} {#3}}}
\nc{\spjetp}[3]{{\it Sov.\ Phys.\ JETP\ }{{\bf #1} {(#2)} {#3}}}
\nc{\yf}[3]{{\it Yad.\ Fiz.\ }{{\bf #1} {(#2)} {#3}}}
\nc{\zetp}[3]{{\it Zh.\ Eksp.\ Teor.\ Fiz.\  }{{\bf #1}  {(#2)} {#3}}}
\nc{\zp}[3]{{\it Z.\ Phys.\ }{{\bf #1} {(#2)} {#3}}}
\nc{\ibid}[3]{{\sl ibid.\ }{{\bf #1} {#2} {#3}}}

\nc{\rf}[1]{(\ref{#1})}
\nc{\nn}{\nonumber \\*}
\nc{\bfB}{\bf{B}}
\nc{\bfv}{\bf{v}}
\nc{\bfx}{\bf{x}}
\nc{\bfy}{\bf{y}}
\nc{\vx}{\vec{x}}
\nc{\vy}{\vec{y}}
\nc{\oB}{\overline{B}}
\nc{\oI}{\overline{I}}
\nc{\oR}{\overline{R}}
\nc{\rar}{\rightarrow}
\nc{\ti}{\times}
\nc{\slsh}{\hskip-5pt/}
\nc{\sm}{Standard~Model~}
\nc{\MP}{M_{\rm Pl}}
\nc{\tp}{t_{\rm Pl}}

\nc{\pmin}{p_{\rm min}}
\nc{\pmax}{p_{\rm max}}
\nc{\fo}{f_0}
\nc{\foi}{f_{0,i}\,}
\nc{\fop}{f_0^P}
\nc{\fou}{f_0^U}

\nc{\eff}{{\rm eff}}
\nc{\MT}{M_{\rm T}}
\nc{\ML}{M_{\rm L}}
\nc{\kk}{\vek{k}}
\nc{\pp}{{\rm p}}
\nc{\pt}{\partial_t}
\nc{\half}{{1\over 2}}
\nc{\w}{\omega}
\nc{\uhat}{\hat{U}_\w}

\nc{\etal}{\mbox{\it et al.}}
\nc{\ie}{{\it i.e. }}
\nc{\eg}{{\it e.g. }}
\nc{\trh}{T_{\rm RH}}
\nc{\ad}{{a'\over a}}
\nc{\bd}{{b'\over b}}
\nc{\Rd}{{R'\over R}}
\nc{\diag}{{\textrm{diag}}}
\nc{\mato}[1]{\tilde{#1}}
\nc{\sech}{\textrm{sech}}
\nc{\I}{\textrm{I}}
\nc{\II}{\textrm{II}}
\nc{\III}{\textrm{III}}
\nc{\vev}[1]{\langle #1 \rangle}

\nc{\brhom}{\overline{\rho}_M}
\nc{\brho}{\overline{\rho}}
\nc{\rhob}{\overline{\rho}}
\nc{\Pb}{\overline{P}}
\nc{\bH}{\overline{H}}

\nc{\hyp}{\,\; F_{1{\hskip -14pt}2}{\hskip 11pt}}
\nc{\lcdm}{$\Lambda$CDM }

\begin{document}
\title{Large scale structure and the generalized Chaplygin gas as dark energy}

\author{Tuomas Multam\"aki$^1$, Marc Manera$^2$, Enrique Gazta\~naga$^2$}
\affiliation{$^1$Departament E.C.M. and C.E.R., Universitat de Barcelona, Diagonal
647, 08028 Barcelona, Spain}
\affiliation{$^2$ Insitut d'Estudis Espacials de Catalunya, IEEC/CSIC, Gran
Capit\'an 2-4, 08034 Barcelona, Spain}
 
\date{}

\begin{abstract}
The growth of large scale structure is studied in a universe
containing both cold dark matter (CDM) and generalized Chaplygin gas (GCg). 
GCg is assumed to contribute only to the background evolution
of the universe while the CDM component collapses and forms
structures.  We present some new analytical as well as numerical results for 
linear and non-linear growth in such model.
The model passes the standard cosmological distance test
without the need of a cosmological constant (LCDM). But 
we find that the scenario is severely constrained by current observations
of large scale structure. Any small deviations of the 
GCg parameters away from the standard Lambda dominated cosmology (LCDM)
produces substantial suppression for the growth of structures.

\end{abstract}


\keywords{cosmology -- cosmic microwave background}

\maketitle

\section{Introduction}

The current cosmological concordance model that is well supported by
the latest microwave background and supernova type Ia observations,
includes both a large repulsive cosmological term and a cold
dark matter (CDM) component. The nature and origin of these
components is yet unknown, leaving room for novel theoretical explanations 
and new cosmological scenarios. In the standard cosmological
scenario the cosmological term, $\Lambda$, is a true constant but,
as is already suggested by the plethora of papers devoted to the
subject, this need not to be the case.
In addition, the difficulty of explaining the existence and 
value of the cosmological constant from fundamental physics, 
leaves plenty of room and motivation for phenomenological proposals.

Instead of having a universe with a non-zero $\Lambda$, one can
study cosmological models that, within the observational
boundaries, imitate the evolution of a 
universe with a cosmological constant and cold dark matter, the
\lcdm model. Such models can have be the result of a modified
Friedmann equation or contain a new type of matter with positive
energy density and negative pressure. Among these models
are the large class of quintessence models (see e.g. \cite{quint}), 
the brane inspired models \cite{dgp}, Cardassian models \cite{car} and
recently the Chaplygin\footnote{Sergei A. Chaplygin, 1869-1942} 
gas models \cite{chap}.

The General Chaplygin gas (GCg) models are characterized by a 
perfect fluid with a non-standard equation of state
\be{chapstate}
p_G=-{\tilde{A}\over\rho_G^\alpha}
\ee
where $\tilde{A}$ is a positive (dimensionfull) constant.
It is useful to define a dimensionless constant, $A$ (also often denoted
as $A_s$), by $A\equiv\rho_{G,0}^{-(1+\alpha)}\tilde{A}$, where
$\rho_{G,0}$ is the present day density of the CGg.

The original Chaplygin gas model \cite{chap} corresponds to the choice $\alpha=1$.
Such an equation of state can be the result of a scalar field with a
non-standard kinetic term, \eg the string theory motivated tachyon
field \cite{sen,gibbons}. The generalization of the original Chaplygin
gas, \ie to cases where $\alpha\neq 1$, was done in \cite{bento}. 
Since the original Chaplygin gas scenario is reached as a special
case of the general Chaplygin gas, we will concentrate on the latter.

The cosmological behavior of the GCg is between a dark matter 
and a cosmological constant. At large densities, it behaves as pressureless
dust where as at small densities, \ie presently, it acts as a cosmological
constant. As such, one can hope to have a unified dark matter
candidate, \ie one that can play both the role of cold dark matter and
of the cosmological constant \cite{chap, makler}. Taking a more
conservative view, one can also consider a universe with both
significant CDM and GCg components.
The different cosmological models that include a (generalized) Chaplygin gas component can 
then be divided into two classes: models with and without a significant CDM 
component. Models that do not include a specific CDM component are often
called unified dark matter (CDM) models.

The unified dark matter models with a (generalized) Chaplygin gas 
have been studied in various works in view of cosmological observations.
Cosmological perturbations of the GCg-fluid have been considered in 
\cite{bento},\cite{fabris}-\cite{avelino}.
In \cite{sandvik} it was claimed that the matter power spectrum
strongly constrains the parameter space of unified dark matter models, 
essentially ruling them out. This result does not include the effect of the
baryons, however, and is hence questionable\cite{carlos}.
Also, as it was pointed out in 
\cite{reis,makler2}, the restriction due to the matter power spectrum
may be avoided by \eg allowing for entropy perturbations. 
To complicate the matter further,
in \cite{avelino} it was shown that one must be careful in
using linear theory to study the perturbations as linear theory may
break down sooner than expected.

Other observational measures of universes with a GCg-component 
include the SNIa constraints \cite{makler,bean,fabris3,colistete,silva,cunha,avelino2},
constraints coming from the CMB \cite{bento,bean,bento2,bento3,amendola,carturan}
and other astrophysical data \cite{makler2,dev,alcaniz}.

A significant fraction of the work so far has concentrated on 
the unified dark matter proposal.
In this letter we concentrate on the more general case where 
there is in addition to the GCg-component, a significant cold dark matter
fraction. The GCg plays the role of dark energy, from this point of view
this model can be seen as a quintessence model with a dynamical
equation of state.
GCg as dark energy has been studied in recent works \cite{cunha,colistete,fabris3,amendola,bean,dev}. Interestingly, in \cite{fabris3, colistete}
it was found by using SNIa data that a universe dominated with a Chaplygin 
gas was favored over a mixed model. On the other hand, in \cite{bean} 
it was concluded that a \lcdm model is preferred. Also in \cite{amendola},
using WMAP data it was showed that a CDM model is disfavored and
a mixed model, with a \lcdm-like behavior, was preferred.
The current bounds on the parameters are varied, depending on the
type observational data. Using globular clusters to determine the age 
of the universe,
in \cite{dev} it was found that $\Omega_M\geq 0.2,\ A\geq 0.96$ ($\alpha=1$), 
where as for the lensing statistics the bounds are somewhat looser
$\Omega_M\leq 0.45,\ A_\geq 0.72$. Using the WMAP data, in \cite{amendola}
the parameter bounds were found to be $A\geq 0.8,\ 0\leq\alpha\leq 0.2\ (95\% CL)$ and the original Chaplygin gas is ruled out.
In \cite{cunha}, X-rays from galaxy clusters, HST data and SNIa data
were used to obtain $A\geq 0.84,\ 0.273\leq \Omega_M\leq 0.329\ (95 \% CL)$.
Using only SNIa data, it was found in \cite{fabris3} for the original
Cg-model that $A=0.93^{+0.07}_{-0.20},\ 0\leq\Omega_M\leq 0.35$.

In this work we study the mixed scenario, with CDM and GCg, from the point of view
of large structure growth. We take a quintessence-like approach to the Chaplygin gas,
\ie only the CDM component is assumed to collapse and form structures. The Chaplygin gas
acts only as dark energy and is visible, from the point of structure growth,
only through the Friedmann equation. This is to be contrasted
with the unified dark matter scenario where the Cg is also responsible for the growth
of structures. The scenario considered here may be seen \eg as a universe dominated 
by a scalar field with a non-standard kinetic term arising from string theory. 
Similar considerations for quintessence scenarios have been presented in \cite{benabed}, 
where large scale structure is used to test different quintessence models.

By studying large scale structure growth from the point of view of gravitational
collapse we probe the significance of the GCg dark energy component throughout the epoch
during which large scale structures grow. The linear and non-linear growth can be related
to observations and hence lead to observational constraints like galaxy number counts
(see \cite{mgm}). Our approach is similar 
to \cite{mgm,lobo}, where other non-standard cosmologies have been 
considered (also see \cite{lue} for related discussions).
In Section 2 we review and study properties of GCg cosmology.
The spherical collapse formalism is reviewed in Section 3 and applied
to the GCg case, producing some analytical results. Numerical work
is performed in Section 4. Discussion and conclusion is presented in
Section 5.

\section{Generalized Chaplygin gas cosmology}

Assuming in general that we have both a cold dark matter
component and a GCg component, the Friedmann equation is 
\be{fried}
H^2\equiv \Big({\dot{a}\over a}\Big)^2=\kappa^2(\rho_{M}+\rho_{G}),
\ee
where $\kappa^2=8\pi G/3$, $\rho_{M}$ is the energy density
of cold dark matter and $\rho_G$ of the generalized Chaplygin gas.
Note that there is no explicit cosmological constant and that we
choose work in a flat universe.

The continuity equation of the different components are
\bea{contequs}
\dot{\rho}_M+3 H(\rho_M+p_M) & = & 0\nonumber\\
\dot{\rho}_G+3 H(\rho_G+p_G) & = & 0,\nonumber
\eea
\ie the two fluids interact only gravitationally. Using Eq. (\ref{chapstate})
and the fact that in a matter dominated universe $p_M=0$, we get
two equations that can be solved straightforwardly:
\bea{contequs2}
\rho_M & = & {\rho_{M,0}\over a^3}\nonumber\\
\rho_G & = & \rho_{G,0}\Big(A+(1-A)a^{-3(1+\alpha)}\Big)^{1\over 1+\alpha},
\eea
where the index $0$ refers to present day values ($a_0=1$) and 
$A(\equiv\rho_{G,0}^{-(1+\alpha)}\tilde{A})$ is a dimensionless
constant, $0\leq A \leq 1$.
From Eq. (\ref{contequs2}), we see that at early times, the GCg
behaves as matter, $\rho_G\sim a^{-3}$, whereas at late times it
imitates a cosmological constant, $\rho_G\approx const.$

Using the expressions (\ref{contequs2}) and the Friedmann
equation, Eq. (\ref{fried}), one can look for solutions of
$a(t)$ or $\rho(t)$. In the general case no analytical solution is found.
As a special case, one sees that in a universe with only 
Chaplygin gas, \ie $\rho_M=0$, the Friedmann equation is integrable
and the solution is (up to a constant):
\be{chapsolua}
t={2\over 3} \Big({1+{A\over 1-A} a^{3(1+\alpha)}\over
A+(1-A) a^{3(1+\alpha)}}\Big)^{1\over 2(1+\alpha)}\,
\hyp({1\over 2(1+\alpha)},{1\over 2(1+\alpha)},1+{1\over 2(1+\alpha)},
{A\over A-1} a^{3(1+\alpha)}).
\ee
In the same special case, one can also solve for the energy
density (again, up to a constant):
\be{chapsolurho}
t=\frac 23 {1\over A\kappa(1+2\alpha)} {\rho^{\frac 12+\alpha}\over\rho_0^{1+\alpha}}\,\hyp({1+2\alpha\over 2(1+\alpha)},1,{3+4\alpha\over 2(1+\alpha)},
{\rho^{1+\alpha}\over A\rho_0^{1+\alpha}})
\ee

Using the solutions (\ref{contequs2}), we rewrite the Friedmann equation in
terms of redshift as
\be{fried2}
H^2=H_0^2\Big(\Omega_M(1+z)^3+\Omega_G(A+(1-A)(1+z)^{3(1+\alpha)})^{1\over 1+\alpha}\Big),
\ee
where we have defined $\Omega_M=\kappa^2\rho_{M,0}/H_0^2$ and
$\Omega_G=\kappa^2\rho_{G,0}/H_0^2$. For $\alpha=0$ we see that 
Eq. (\ref{fried2}) reduces to the \lcdm-model with 
$\Omega_m=\Omega_M+\Omega_G(1-A)$ and $\Omega_{\Lambda}=\Omega_G A$.
Note that in a flat universe, our definitions imply that $\Omega_M+\Omega_G=1$.

%

\begin{figure}[ht]
\center
\includegraphics[width=90mm,angle=-90]{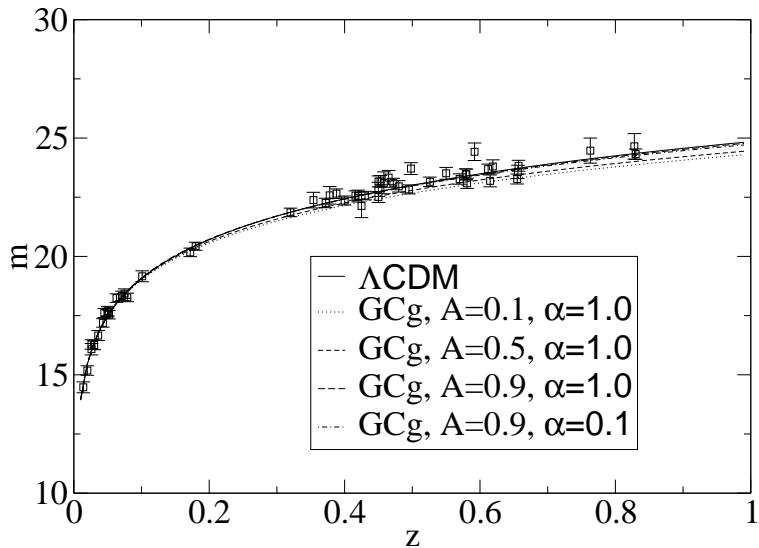}~~
\caption{Apparent magnitudes for SNIa for different cosmologies ($\Omega_M=0.3,\ \Omega_G=0.7$).}
\label{sniafig} 
\end{figure}

As a first test, we should compare distances in the GCg cosmology 
with the standard \lcdm case. 
The luminosity distance $d_L$ is obtained from the 
(line-of-sight) comoving coordinate distance: $r(z)=\int dz'/H(z')$.
Fig. \ref{sniafig} shows predictions for different values of the model
parameters for the corresponding apparent magnitude 
as compared to the SNIa results \cite{rei,per}.
It is apparent that 
current  observations do not discriminate much between the different 
cosmological models. In other words: 
the GCg cosmologies seem to pass the standard cosmological distance test.
It is then interesting to explore whether they also pass the observational 
constraints on the growth of linear and non-linear structures.

\section{Gravitational collapse}

The gravitational evolution of an over-dense region is dependent
on the background evolution. A rapidly expanding background
will obviously slow down the collapse of region compared to a
static background. The details of the collapse are obviously
complicated in a general case so here we choose work in a
shear-free approximation and consider spherically symmetric
configurations, \ie we use the spherical collapse model to study
the growth of large scale structure.

To study the dynamics of gravitational collapse, it is useful
to define local density contrast as
\be{ldc}
\delta={\rho\over\bar{\rho}}-1,
\ee
where $\bar{\rho}$ is the background energy density.
Within the spherical collapse approximation one can show that
in a matter dominated universe (\ie the collapsing matter has no pressure)
\cite{mgm,lobo}
\be{gendenspe}
{d^2\delta\over d\eta^2}+(2+{\dot{\bH}\over \bH^2}){d\delta\over d\eta}
-\frac 43{1\over 1+\delta}({d\delta\over
d\tau})^2=-3{1+\delta\over\bH^2}\Big((\dot{H}+H^2)-(\dot{\bH}+\bH^2)\Big),
\ee
where $\eta=\ln(a)$, $\bH=\bH(\bar{\rho})$ is the background Hubble rate
and $H=H(\delta,\bar{\rho})$ is the local Hubble rate.

Working within the framework of small perturbations we expand $\delta$
as
\be{deltaexp}
\delta=\sum_{i=1}^{\infty}\delta_i=\sum_{i=1}^{\infty}{D_i(\eta)\over i!}\delta_0^i,
\ee
where $\delta_0$ is the small perturbation.
Similarly, we can also expand the $\dot{H}+H^2$-term in terms of 
$\delta$ and then 
the whole RHS of Eq. (\ref{gendenspe}) as
\be{genrhsexp}
3{1+\delta\over\bH^2}\Big((\dot{H}+H^2)-(\dot{\bH}+\bH^2)\Big)
\equiv 3(1+\delta)\sum_{n=1} c_n\delta^n.
\ee
Using this and expanding the perturbation according to Eq. (\ref{deltaexp}) 
we find the linear equation,
\be{genlinear}
D_1''+(2+{\dot{\bH}\over \bH^2})D_1'+3 c_1D_1=0,
\ee
the second order equation,
\be{gen2nd}
D_2''+(2+{\dot{\bH}\over \bH^2})D_2'-\frac 83 (D_1')^2+3c_1D_2
+6(c_1+c_2)D_1^2=0,
\ee
and one can go on to arbitrary order by using the solutions to
the lower order equations.

The second order equation can be related to the skewness of the
density field at large scales  (see \cite{bcgs} for more details). 
For Gaussian initial conditions,
\be{s3}
S_3=3 {D_2\over D_1^2}.
\ee

In an Einstein-deSitter universe, 
\bea{eds}
c_1 & = & -\frac 12\\
c_i & = & 0,\ i=2,3,...\\
2+{\dot{\bar{H}}\over \bar{H}^2} & = & \frac 12
\eea
and the well known solution to the linear equation (\ref{genlinear})
is 
\be{edssol}
D_1=B_1e^\eta+B_2e^{-\frac 32\eta}=B_1a+B_2a^{-\frac 32}.
\ee
The skewness can also be calculated analytically (see \cite{bcgs}):
\be{edss3}
S_3^{EdS}={34\over 7}\approx 4.86.
\ee

\subsection{Gravitational collapse in GCg}

The $\dot{H}+H^2$-term in the GCg model is
\be{doth}
\dot{H}+H^2=-\frac 12\kappa^2\Big(\rho_M+\rho_G(1-3
{\tilde{A}\over \rho_G^{1+\alpha}})\Big)
\ee

The gravitational dynamics in the GCg model obviously depend on
the type of scenario \ie whether we have a separate dark matter 
component or not and which components are fluctuating to form structures.


Assuming that GCg only affects large scale structure growth through
its effect on the background evolution, \ie only the fluctuations of the CDM
component are responsible for large scale structure, we can write
$\rho_M=\bar\rho_M (1+\delta)$, $\rho_G=\bar\rho_G$. 
Expanding according to (\ref{genrhsexp}) we find that 
\bea{decs}
c_1 & = & -\frac 12 {\bar\rho_M\over\bar\rho_M+\bar\rho_G} = 
-\frac 12
{\Omega_M\over\Omega_M+\Omega_G[1+A(a^{3(1+\alpha)}-1)]^{1\over 1+\alpha}}
\\
c_2 & = & 0\\
2+{\dot{\bar{H}}\over\bar{H}^2} & = & \frac 12 (1+3
{\tilde{A}\over\bar\rho_G^{\alpha}(\bar\rho_M+\bar\rho_G)})
 =  \frac 12\Big(1+{3 A \Omega_G [A+(1-A)a^{-3(1+\alpha)}]^{-{\alpha\over
1+\alpha}}\over\Omega_M a^{-3}+\Omega_G
[A+(1-A)a^{-3(1+\alpha)}]^{1\over 1+\alpha}}\Big)
\eea

From Eqs (\ref{decs}) the small $a$ limit, \ie the
early time limit, is easily read:
\bea{decslim1}
c_1 & = & -\frac 12 {\Omega_M\over\Omega_M+\Omega_G(1-A)^{1\over
1+\alpha}}\\
2+{\dot{\bar{H}}\over\bar{H}^2} & = & \frac 12.
\eea
Comparing this to the linear equation in a \lcdm-universe,
\be{lcdm}
D_1''+\frac 12 {\Omega_M+4 a^3\Omega_{\Lambda}\over\Omega_M+a^3\Omega_{\Lambda}}D_1'-\frac 32
{\Omega_M\over\Omega_M+a^3\Omega_{\Lambda}}=0,
\ee
we see that a GCg-universe behaves fundamentally differently from the
\lcdm-universe:
at early times, $a\ll 1$, \lcdm-universe reduces to the
EdS-case, with solution given by Eq. (\ref{edssol}),
where as the GCg-universe does not. Accordingly, the solution to the linear
equation in the GCg-case (in the small $a$ limit) is
\be{GCgsolu}
D_1=B_1 e^{\frac 14 (\sqrt{1+24\xi}-1)\eta}+B_2 e^{\frac 14 (-\sqrt{1+24\xi}-1)\eta},
\ee
where
\be{xidef}
\xi={\Omega_M\over\Omega_M+\Omega_G(1-A)^{1\over 1+\alpha}}.
\ee
Hence, EdS-case is recovered in the limits $\Omega_G=0$ and $A=1$.
Compared to the EdS-case (and \lcdm-universe), fluctuations
therefore start to grow more slowly in a universe where GCg plays a
role. Defining $\epsilon=(1-A)^{1/(1+\alpha)}\Omega_G/\Omega_M$, we see
that the growing mode goes as (when $\epsilon\ll 1$)
\be{GCggrow}
D_1\sim a^{1-\frac 35 \epsilon}.
\ee

On the other hand, in the large $a$ limit,
\be{GCgbiga}
D_1''+\frac 12 {\Omega_M+4\Omega_G A^{1\over
1+\alpha}a^3\over\Omega_M+\Omega_G A^{1\over 1+\alpha}a^3}
D_1'-\frac 32 {\Omega_M\over\Omega_M+\Omega_G A^{1\over
1+\alpha}a^3}D_1=0,
\ee
which is the standard \lcdm equation, Eq. (\ref{lcdm}),
with $\Omega_{\Lambda}=\Omega_G A^{1\over 1+\alpha}$.

The second order equation reads at early times as
\be{GCg2nd}
D_2''+\frac 12 D_2'-\frac 32 \xi D_2-\frac 83(D_1')^2-3\xi D_1^2=0.
\ee
Substituting the growing mode from Eq. (\ref{GCgsolu}), the solution to
this equation is
\be{GCg2ndsol}
D_2=\frac 43 {\sqrt{1+24\xi}-1-21\xi\over\sqrt{1+24\xi}-1-18\xi}
e^{\frac 12(\sqrt{1+24\xi}-1)\eta}+B_3 e^{\frac
14(\sqrt{1+24\xi}-1)\eta}+B_4 e^{-\frac 14(\sqrt{1+24\xi}-1)\eta}.
\ee
Hence, at early times, the skewness is given by
\be{s3new}
S_3=3 {D_2\over D_1^2}=4
{\sqrt{1+24\xi}-1-21\xi\over\sqrt{1+24\xi}-1-18\xi},
\ee
which when $\Omega_G=0$ (\ie $\xi=1$) reduces to the EdS-value, 
$S_3^{EdS}=34/7$.
At the other extremum point, $\Omega_M=0$ or $\xi=0$, we reach 
$S_3=6$. All the other values lie between these two limits.
Numerical work (see later section) shows that Eq. (\ref{s3new})
is a very good approximation until very late times, but even there
the error is very small.


\begin{figure}[ht]
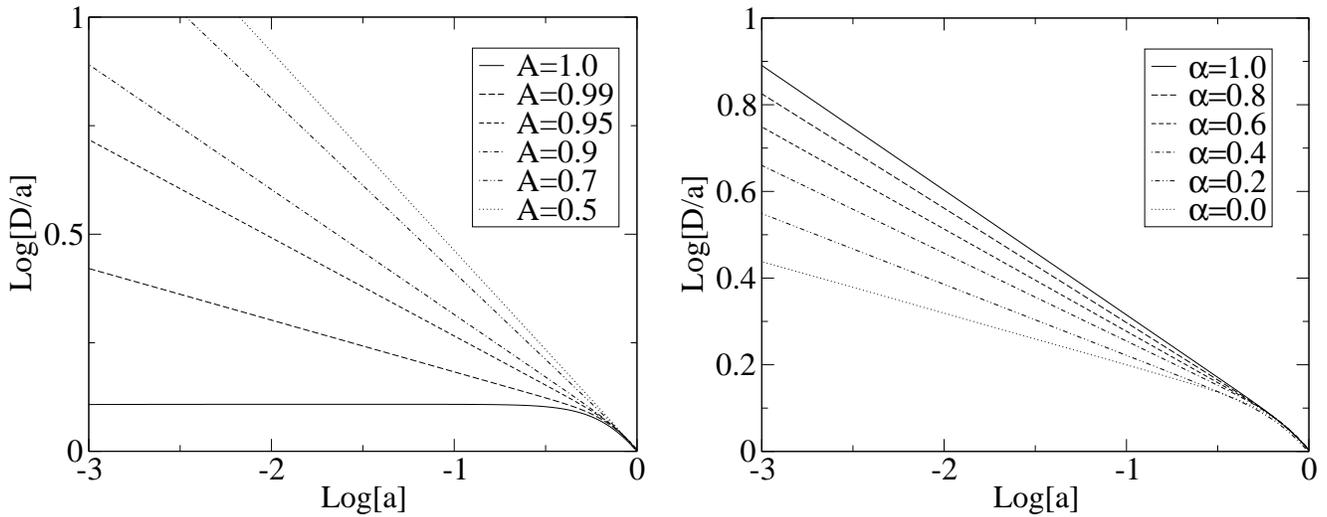

\includegraphics[width=85mm,angle=0]{grafD_A.eps}~~~
\includegraphics[width=85mm,angle=0]{grafD_al.eps}
\caption{Linear growth for $\alpha=1.0$, $A=1, 0.95, 0.9, 0.7, 0.5$
(left panel), $A=0.9$, $\alpha=0, 0.2,..., 1.0$(right panel)}
\label{linfig}
\end{figure}     

\subsection{Numerical Results}

In order to study the linear and non-linear evolution of the
perturbation from early to late times, we must resort to
numerical means. The linear and second order equations are solved
numerically over the region of cosmological interest, from $a=10^{-3}$
to $a=1\equiv a_0$. The initial conditions are chosen such that in both cases,
the growing part of the solution, Eqs (\ref{GCgsolu}) and
(\ref{GCg2ndsol}), is followed at early times.
This ensures that the
behavior of the solutions is qualitatively similar to the
\lcdm-universe and that the \lcdm-behavior is reached
as $A$ tends to unity. Normalization is chosen so that
the amplitude of fluctuations is the same at present time, \ie at $a=1$.

The linear factor is are plotted in Fig \ref{linfig}. Note that 
the linear growth is depicted relative to $a$, \ie to the growth rate in the EdS-universe (which is also equal to the growth rate in  \lcdm-universe at early times). In plotting the figures, we choose $\Omega_M=0.3$, which in a flat
universe sets $\Omega_G=0.7$. 

The linear growth rate confirms what was expected from the analytical
results: linear growth occurs much more slowly than in a EdS
(\lcdm) universe until at late times when the GCg acts as a
cosmological constant that dominates, leading to even stronger
suppression of growth. Note that the case $A=1.0$ corresponds to the
\lcdm universe with $\Omega_M=0.3, \Omega_{\Lambda}=0.7$.
The effect of changing $\alpha$ is much less significant as is shown
in Fig. \ref{linfig} (right panel). 

\begin{figure}[ht]
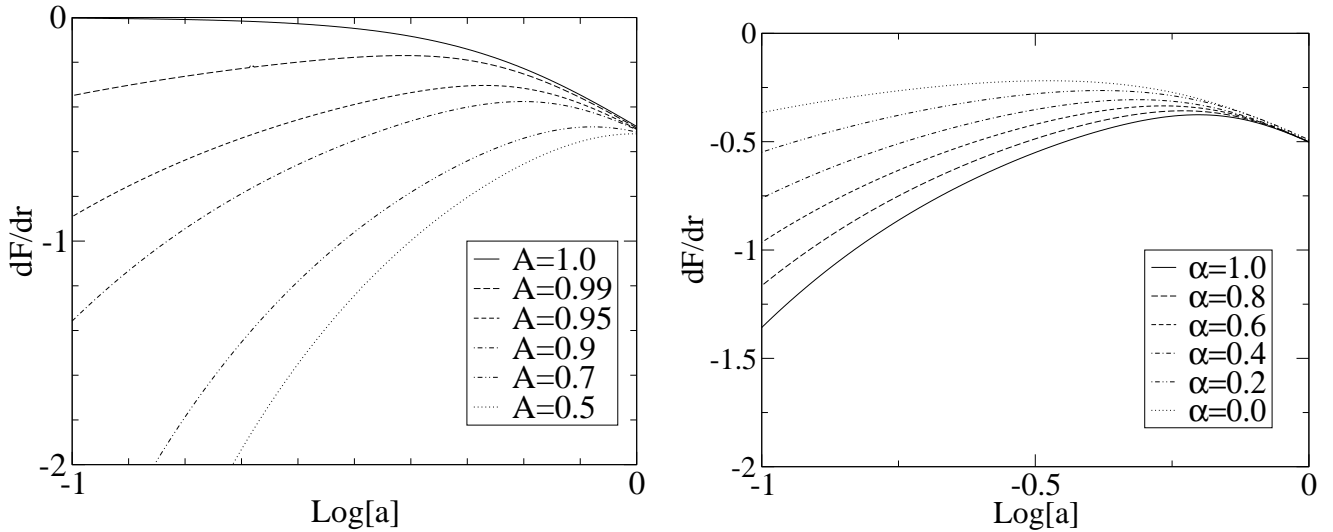

\includegraphics[width=85mm,angle=0]{grafF_A.eps}~~~
\includegraphics[width=85mm,angle=0]{grafF_al.eps}
\caption{Derivative of the linear growth: $\dot{F}=d(D/a)/dr$
 for $\alpha=1.0$, $A=1, 0.95, 0.9, 0.7, 0.5$
(left panel), $A=0.9$, $\alpha=0, 0.2,..., 1.0$(right panel)}
\label{fdotfig}
\end{figure}

It is also interesting to study the time derivative of the above
ratio: $dF/dr=d(D/a)/dr$, where $dr$ is the comoving distance.
This quantity provides a direct observable (see eg
\cite{Fosalba} and references therein)
through the integrated Sachs-Wolfe (ISW) effect
\cite{SW,CT96}. Numerical predictions for $dF/dr$ are shown
in Fig. \ref{fdotfig}. As expected deviations from the
standard \lcdm model are substantial for most of the relevant
GCg parameters.

The effect on the skewness is not as dramatic
and is very well approximated by the analytical formula Eq.(\ref{s3new})

Further details on how these quantities can be compared to observations,
can be found in \cite{lobo,mgm} and references therein.

\section{Discussion}

We have considered a simple variation on  the Generalized Chaplygin gas
cosmology, where  in addition to the GCg-component there is
a significant cold dark matter fraction. The GCg plays the 
role of dark energy, but not in a unified way (see Introduction) meaning that
the GCg does not collapse gravitationally.
At first look this model seem viable, as it has quite a standard
looking Friedmann equation, see  Eq. (\ref{fried}), and cosmological 
distances, see  Fig. \ref{sniafig}. However, this is not the case
for the growth of structures.

As we have seen both analytically and numerically, the growth of large
scale structure is fundamentally different from the \lcdm scenario due
to the properties of the Generalized Chaplygin gas. Unlike in the
\lcdm-model, where at early times the equation governing linear growth
equals that of the EdS-case, the GCg behaves differently at early
times. From Eq. (\ref{GCggrow}) we see that for $A\neq 1,\ \rho_{G,0}\neq 0$, 
linear growth will be suppressed relative to the $EdS$ scenario, and
to a \lcdm universe which follows the EdS-solution until late times.
We can see in  Fig. \ref{linfig} that any small deviations of the 
GCg parameters from the \lcdm values ($\alpha=1$ and $A=1$) produced
substantial suppression of linear growth. For example, even the case
$A=0.99$ produces a much higher amplitude of fluctuations
at decoupling ($a \simeq 10^{-3}$) than the \lcdm cosmology (when normalized
at $a=1$). Such differences in normalization are hard to reconcile
with our current understanding of CMB anisotropies (see eg \cite{mgm,spergel}).
The final numbers dependent on the adopted model for CMB transfer function
and matter content, but it is already apparent that we will need to
tune the GCg parameters very close to the \lcdm model. As an independent test
we have also presented the predictions of the GCg model for the ISW effect,
see Fig.\ref{fdotfig} and the skewness of the density field, see Eq.(\ref{s3new}).
Again, the ISW effect is much larger even for $A=0.99$ compared to 
a \lcdm universe. It should be stressed that in the scenario considered here,
the GCg only acts through the background evolution and does not undergo gravitational collapse.
Hence, these parameter constraints do not apply to the unified dark matter models with a 
GCg component.

Observations seem to favor values of  the GCg parameter which 
makes it equivalent in practice to a cosmological constant.
We conclude therefore that using the GCg as a form of dark energy does
not seem to provide any apparent advantage over the more simple \lcdm model,
which has fewer parameters. It is nevertheless clear that future observations will be able 
to provide tighter constraints on the quantities we have studied here. If the \lcdm model fails to match
such observations, the GCg models could be an useful alternative to consider.

\acknowledgments
TM is grateful to the Academy of Finland (grant no. 79447) for
financial support and to the Theoretical Physics department of Oxford
University for hospitality during the completion of this work.
EG and MM acknowledge support from
and by grants from IEEC/CSIC and the spanish Ministerio de Ciencia y
Tecnologia, project AYA2002-00850 and EC FEDER funding.
MM acknowledges support from a PhD grant from Departament d'Universitats,
Recerca i Societat de la Informacio de la 
Generalitat de Catalunya.
We are grateful to the 
Centre Especial de Recerca en Astrofisica, Fisica de Particules i Cosmologia
(C.E.R.) de la Universitat de Barcelona and IEEC/CSIC 
for their support.


\end{document}